\documentclass[lettersize,journal]{IEEEtran}
\usepackage[dvipsnames]{xcolor}
\usepackage{amsmath,amsfonts}
\usepackage{algorithmic}
\usepackage{algorithm}
\usepackage{array}
\usepackage{textcomp}
\usepackage{stfloats}
\usepackage{url}
\usepackage{verbatim}
\usepackage{graphicx}
\usepackage{cite}

\usepackage{booktabs}
\usepackage{multirow}

\usepackage{longtable}

\usepackage{float}
\usepackage{subfigure}
\usepackage{balance}

\usepackage{xcolor}
\usepackage{soul}

\usepackage[shortcuts,acronym]{glossaries}

\newacronym{thz}{THz}{Terahertz}
\newacronym{los}{LOS}{line-of-sight}
\newacronym{6g}{6G}{sixth-generation}
\newacronym{kl}{KL-divergence}{Kullback-Leibler Divergence}
\newacronym{pdfs}{PDFs}{probability density functions}
\newacronym{mgfs}{MGFs}{moment generation functions}
\newacronym{snr}{SNR}{signal-to-noise ratio}
\newacronym{cdf}{CDF}{cumulative distribution function}
\newacronym{dir}{$Dir$}{Dirichlet distribution}

\newacronym{em}{EM}{expectation-maximization}
\newacronym{dp}{DP}{Dirichlet process}
\newacronym{gmm}{GMM}{Gamma mixture model}
\newacronym{dpm}{DPM}{Dirichlet process mixture}
\newacronym{dpmm}{DPMM}{Dirichlet process mixture model}
\newacronym{dpgm}{DPGM}{Dirichlet process gamma mixture}
\newacronym{dpgmm}{DPGMM}{Dirichlet process Gamma mixture model}

\newacronym{bic}{BIC}{Bayesian information criteria}

\newacronym{mcmc}{MCMC}{Markov Chain Monte Carlo}
\newacronym{nuts}{NUTS}{No-U-Turn Sampler}
\newacronym{hmc}{HMC}{Hamiltonian Monte Carlo}

\renewcommand{\arraystretch}{0}

\begin{document}

\title{Hierarchical Dirichlet Process Based Gamma Mixture Modelling for Terahertz Band Wireless Communication Channels}
\author{Erhan Karakoca,~\IEEEmembership{Graduate Student Member,~IEEE,} G\"{u}ne\c{s} Karabulut Kurt~\IEEEmembership{Senior Member,~IEEE,} \newline Ali G\"{o}r\c{c}in~\IEEEmembership{Senior Member,~IEEE}

\thanks{Erhan Karakoca is with the Department of Electronics, and Communications Engineering, \.Istanbul Technical University, 34467 Istanbul, Turkey, and also with HISAR Laboratory, Informatics and Information Security Research Center (B\.ILGEM), T\"UB\.ITAK, 41470 Kocaeli, Turkey (e-mail: karakoca19@itu.edu.tr).}
\thanks{G. Karabulut Kurt is with the Poly-Grames Research Center, Department of Electrical Engineering  Polytechnique Montr\'eal, Montr\'eal, Canada (e-mail: gunes.kurt@polymtl.ca).}
\thanks{Ali G\"{o}r\c{c}in is with HISAR Laboratory, Informatics and Information Security Research Center (B\.ILGEM), T\"UB\.ITAK, 41470 Kocaeli, Turkey, and also with Department of Electronics and Communications Engineering, Y{\i}ld{\i}z Technical University, 34349 Istanbul, Turkey (e-mail: agorcin@yildiz.edu.tr).}
}

\markboth{}%
{Shell \MakeLowercase{\textit{et al.}}: A Sample Article Using IEEEtran.cls for IEEE Journals}


\maketitle

\begin{abstract}
Due to the unique channel characteristics of \ac{thz},  comprehensive propagation channel modeling is essential to understand the spectrum and develop reliable communication systems in these bands. 
In this work, we propose the utilization of the hierarchical \ac{dpgmm} to characterize \ac{thz} channels statistically in the absence of any prior knowledge.
\ac{dpgmm} provides mixture component parameters and the required number of components. 
A revised \ac{em} algorithm is also proposed as a pre-step for \ac{dpgmm}.
\ac{kl} is utilized as an error metric to examine the amount of inaccuracy of the \ac{em} algorithm and \ac{dpgmm} when modeling the experimental \ac{pdfs}. 
\ac{dpgmm} and \ac{em} algorithm are implemented over the measurements taken at frequencies between 240 GHz and 300 GHz. 
By comparing the results of the \ac{dpgmm} and \ac{em} algorithms for the measurement datasets, we demonstrate how well the \ac{dpgmm} fits the target distribution. 
It is shown that the proposed \ac{dpgmm} can accurately describe the various \ac{thz} channels as well as the \ac{em} algorithm, and its flexibility allows it to represent more complex distributions better than the \ac{em} algorithm. 
We also demonstrated that \ac{dpgmm} can be used to model any wireless channel due to its versatility. 
\end{abstract}


\section{Introduction}
\IEEEPARstart
{T}{he} demand for high-bandwidth instant online connectivity grows every day.
That is prompting the emergence of new data-hungry technological instruments, resulting in an everlasting increase in wireless data traffic load \cite{cherry2004edholm, cisco2020cisco}.
As a result, both academia and the industry increased their interest in higher frequency bands featuring wider bandwidths to comply with the demand. 
The \acrfull{thz} band between 0.1 \ac{thz} and 10 \ac{thz} is one of the spectra that has been examined from various perspectives recently \cite{akyildiz2014terahertz,tekbiyik2019terahertz} and appears to have a promising future.
Since the \ac{thz} band is not standardized and allows bandwidths in the orders of 100 GHz, researchers push for the design of \ac{thz} wireless systems that will enable communication~\cite{tekbiyik2020holistic}.
Owing to its appealing features, the \ac{thz} band is expected to play a key role in \ac{6g} communication systems by providing data rates up to Terabit per second levels \cite{yang20196g}.

Besides these, \ac{thz} band comes at the cost of severe losses such as propagation losses and molecular absorption.
Because of these issues, \ac{thz} band has distinct channel characteristics than other commonly utilized frequency bands. Even various \ac{thz} sub-bands have different channel characteristics, demonstrating the importance of channel modeling in \ac{thz} band and the requirement for a range of flexible channel modeling approaches.
\subsection{Related Works}
Several works employing various approaches have been conducted to achieve the goal of accurate channel modeling in \ac{thz} bands.
Generally, channel models are divided into two types \textit{i.e.}, deterministic and statistical. 

In terms of deterministic channel modeling, ray-tracing is frequently utilized.
In \cite{priebe2013ultra}, the frequency domain ray-tracing approach is employed to estimate the \ac{thz} indoor propagation channel based on the measurements conducted at 275 to 325 GHz. 
In \cite{guan2021channel}, the smart rail mobility channel is characterized by ray-tracing at the 300 GHz band.
Ray-tracing based on a stochastic channel model of high data rate data download at  220 GHz and 340 GHz is proposed in \cite{he2017stochastic}.
Deterministic modeling methods provide accurate channel modeling results; however, the detailed geometric structure of the propagation environment, as well as the transmitter and receiver locations must be known beforehand, which becomes a significant problem, especially in case of mobility. 
Furthermore, deterministic approaches suffer from high computational complexity.

On the other hand, statistical modeling approaches can be employed to estimate the \ac{thz} propagation characteristics.
A statistical path loss model for 240 to 300 GHz band is proposed in \cite{ekti2017statistical}. Furthermore, a two-slope statistical path loss model for short-range \ac{thz} communication links between 275 and 325 GHz based on real measurement results is introduced in \cite{tekbiyik2019statistical}. 
The suitability of the $\alpha-\mu$ distribution for measurements taken at the different transmitter and receiver configurations in a shopping mall, an airport check-in area, and a university entrance hall was investigated in \cite{papasotiriou2021experimentally} to model \ac{thz} small scale fading accurately. 
In addition to these works, the ergodic capacity of \ac{thz} wireless systems is evaluated in \cite{papasotiriou2021new} by defining \ac{thz} wireless channels with a $\alpha-\mu$ distribution.
Statistical modeling techniques are able to characterize \ac{thz} channels in different environments rather than assuming a specific place.
Beyond this, the fundamental advantage of statistical channel modeling is the lower complexity when compared to deterministic models, which enables quick channel model generation based on essential characteristics \cite{han2018propagation}.

Contemporary \ac{thz} band statistical channel modeling approaches mostly assume that the channel can be characterized based on a single distributional representation. 
However, taking into account the variable channel characteristics of \ac{thz} sub-bands, mixed models should be considered. In fact, although the use of mixture models in density estimation is common~\cite{escobar1995bayesian,mclachlan2019finite,rasmussen1999infinite, roeder1997practical,ferguson1983bayesian}, it is rarely used in studies to model wireless communication channels. For example,~\cite{atapattu2011mixture} employed a mixture of Gamma distributions to model the \ac{snr} of various wireless channels by matching \acrfull{pdfs} and \ac{mgfs}, and it enables the evaluation of channel capacity, outage, and error rate owing to its mathematically tractable form and high accuracy representation. Also, in \cite{atapattu2011mixture}, the required number of mixture components is determined by checking values of mean squared error and \acrfull{kl} with the increasing number of mixture components. 
In \cite{buyukccorak2014lognormal}, shadow fading of the empirical results collected in real scenario modeled with lognormal mixtures.
In \cite{selim2015modeling},  the Gaussian mixture model is used to describe wireless channels, where the component parameters and the number of required components are determined by \ac{em} and \ac{bic}, respectively.
When \ac{thz} bands are investigated,  \cite{tekbiyik2021modeling} proposed \ac{gmm} for 240 to 300 GHz band characteristics and represented the bands accurately. In this work, \acrfull{em} algorithm is utilized to infer the parameters of Gamma distribution components.
 
\subsection{Motivation}
\ac{gmm} provided a significant step forward for \ac{thz} channel characterization, however there are 
major improvement points in terms of applications since matching \ac{pdfs} and \ac{mgfs} requires special parametrization for every distribution. 
Moreover, the performance of the \ac{em} algorithm is highly dependent on initialization parameters \cite{meila2013experimental,karlis2003choosing} thus, it has to be tuned every time with the required number of mixture components to be provided a priori. 
To that end, \ac{bic} can be utilized to determine the number of required mixture components. However, in such a scenario, \ac{bic} should scan all potential mixture component numbers to one with the minimum \ac{bic}. This additional step should be repeated for each different distribution. To overcome the impracticality of this process, there a more generalized method is strived for.
Motivated by this requirement, in this work, the utilization of a \ac{dpmm} for \ac{thz} band small scale fading modeling is proposed.
\ac{dpmm} is widely used in many field involving nonparametric bayesian estimation problems such as classification of brain tissues\cite{da2007dirichlet}, protein fold\cite{shahbaba2009nonlinear}, hyperspectral images\cite{wu2016dirichlet} and clustering of genes\cite{dahl2006model}, texts\cite{yin2014dirichlet}, medical images\cite{ elguebaly2015hierarchical}
and density estimation\cite{wiper2001mixtures} of the relevant data in many fields.
\ac{dpmm} has the ability to grow its representation as more data are observed assuming that data come from distinct sub-clusters, with each sub-cluster's data described by a separate probability density function.
Without any prior knowledge, it allows inferring the required number of clusters and statistical properties of each corresponding clusters as data observed.
Because of these properties, \ac{dpmm}s are accurate and adaptable models, especially when the underlying distribution of the data is either known, or can be closely approximated by the assumed model.
As a result of this, \ac{dpmm} has gained considerable popularity in the field of machine learning because of its flexibility, especially in unsupervised learning, due to its  clustering and latent feature extraction property.

The one important thing that has to be considered is clusters in actual data do not always possess symmetric distribution like the Gaussian distribution.  
The Gamma distribution is a versatile alternative to the Gaussian distribution, and because of its skewness, it can describe both long-tailed and asymmetric distributions.
As a result, any arbitrary PDF can be modeled by utilizing a Gamma mixture \cite{wiper2001mixtures}. 
Because of this, Gamma kernels are preferred in \ac{dpmm} studies such as\cite{wiper2001mixtures,copsey2003bayesian,hanson2006modeling,elguebaly2015hierarchical}, to model the density of the real data and their belonging clusters accurately. 
As previously stated, due to its variety, also \ac{thz} sub-bands do not always have the same form and rarely follow a smooth distribution.
For these reasons, Gamma distribution is preferred as the mixture kernel in our proposed model, and we denote it as the \acrfull{dpgmm}.
Unlike the previous mixture-based channel modeling works, distribution parameters and the number of components required for different channel types can be inferred using a single \ac{dpgmm} without any further processing and prior knowledge. 
This allows a single \ac{dpgmm} to be used to describe various distributions from simple to complex.
\subsection{Contributions}
 \begin{itemize}
     \item Motivated by the requirements described, \ac{em} algorithm is revised in this work for \ac{gmm} and as a follow up to the works of~\cite{atapattu2011mixture, tekbiyik2021modeling, buyukccorak2014lognormal} we proposed a flexible hierarchical \ac{dpgmm} for the \ac{thz} band, which determines the required number of mixture components and corresponding mixture component parameters according to the variable structure of \ac{thz} channels.
     We clarified  our contributions on Table \ref{tab:contributions}.
     \item To demonstrate the validity of the models, the proposed \ac{dpgmm} and \ac{em} algorithm are applied to several measurements conducted between 240 and 300 GHz. 
    
     \item The ability of the \ac{dpgmm} and \ac{em} algorithm to estimate experimental distributions is assessed by the \ac{kl} error metric, and results are illustrated. Furthermore, an extensive comparison of the \ac{dpgmm} and \ac{em} algorithm for the \ac{gmm} is carried out.

     \item The flexibility of \ac{dpgmm} is validated, and its ability to span a broad range on the positive axis is proven by applying it to a very large measurement result.
     Therefore, it is shown that \ac{dpgmm} is not only applicable to \ac{thz} channels, but it can also be used to characterize a wide variety of wireless communications channels, including those with single distribution characteristics. Finally, the python code for the developed model is also made available\footnote{DPGMM source codes available to reproduce the results and further research activities: https://github.com/erhankarakoca/DPGMM-Channel-Modelling} for further research activities in this domain.
 \end{itemize}

\begin{table*}[ht!]
\renewcommand{\arraystretch}{1.4}
\setlength{\tabcolsep}{2pt}
\caption{Distinguishing features and contributions}
\label{tab:contributions}
\centering
\begin{tabular}{|c|l|l|l|l|l|}
\hline
\textbf{Reference} & \multicolumn{1}{c|}{\textbf{Subject}}                                                                                                                 & \multicolumn{1}{c|}{\textbf{Approach}}                                                                 & \multicolumn{1}{c|}{\textbf{Mixtures}}                            & \multicolumn{1}{c|}{\textbf{Prior knowledge}} & \multicolumn{1}{c|}{\textbf{Band of interest}}                      \\ \hline
\textbf{{[}20{]}}  & \begin{tabular}[c]{@{}l@{}}Approximation of SNR distribution \\ for any fading channel\end{tabular}                                                   & \begin{tabular}[c]{@{}l@{}}PDF, MGF and moment matching, \\ Bayesian information criteria\end{tabular} & Gamma                                                             & required                                      & \begin{tabular}[c]{@{}l@{}}no bandwidth \\ restriction\end{tabular} \\ \hline
\textbf{{[}21{]}}  & Modeling of shadow fading                                                                                                                             & EM and DPMM                                                                                                 & \begin{tabular}[c]{@{}l@{}}Lognormal and \\ Gaussian\end{tabular} & not required                                  & \textless{}6GHz                                                     \\ \hline
\textbf{{[}23{]}}  & Modeling of short distance THz channel                                                                                                                 & Maximum likelihood and EM                                                                                 & Gamma                                                             & required                                      & 200 - 300 GHz                                                       \\ \hline
\textbf{Ours}      & \begin{tabular}[c]{@{}l@{}}Accurate modeling of THz band channel\\and giving an approach capable of model \\ any of wireless communication channel\end{tabular} & Hierarchical DPMM                                                                                                & Gamma                                                             & not required                                  & \begin{tabular}[c]{@{}l@{}}no bandwidth \\ restriction\end{tabular} \\ \hline
\end{tabular}
\end{table*}
 
\subsection{Organization of the Paper}
This paper is organized as follows.
Section \ref{sec:background} describes the mathematical foundations of the signal model and the \ac{gmm}, as well as the expectation maximization algorithm for the \ac{gmm}.
Section \ref{sec:dpgmm} firstly explains the \ac{dp} and its construction with the stick-breaking process. 
Then introduces the \ac{dpgmm} by giving detailed information about its hierarchical structure.
Section \ref{sec:simulations_and_results} provides detailed information about the measurement dataset and simulation settings. 
Following that,  the performance of the \ac{dpgmm} and \ac{em} algorithms on the measurement dataset is shown, and both models are contrasted.
Finally, Section V concludes the study.

\section{Preliminaries}
\label{sec:background}
\subsection{Signal Model}
    The signal collected at the receiver can be expressed in time domain as
    \begin{equation}
        r(t)=\mathfrak{Re}\left\{\left[x_{I}(t)+j x_{Q}(t)\right] e^{j 2 \pi f_{c} t}\right\},
    \end{equation}
    with $f_c$ being the carrier frequency of the transmitted signal, $j$ representing imaginary number $\sqrt{-1}$. $x_I(t)$ and $x_Q(t)$ define the in-phase and quadrature parts of the received signal, respectively. 
    $\mathfrak{Re}\{\cdot\}$ represents the real-valued terms of the complex baseband signal $r(t)$.
    
    The impulse response of the multipath channel at the passband can be given as
    \begin{equation}
        h(t)=\sum_{l=0}^{L-1} a_{l} \delta\left(t-t_{l}\right),
    \end{equation}
    where $L$ is the total number of multipath sources and $\delta$ is the Dirac delta function. $a_l$ and $t_l$ denote attenuation and delay coefficients for the given $l^{th}$ multipath source. 
    The corresponding complex baseband representation of the multipath channel impulse response is
    \begin{equation}
        h(t)=\sum_{l=0}^{L-1} a_{l} \delta\left(t-t_{l}\right) e^{-j 2 \pi f_{c} t_{l}}.
        \label{imp_resp_mult}
    \end{equation}
    In the case of only the \ac{los} component, $L=1$ in Eq. \eqref{imp_resp_mult} and \ac{los} channel can be given as
    \begin{equation}
        h(t)=a_{0} \delta\left(t-t_{0}\right) e^{-j 2 \pi f_{c} t_{0}},
    \end{equation}
    where $a_0$ is amplitude and $2\pi f_ct$ phase of the channel. 
    Propagation delay; $t_0 =d/c$, where d is the spacing between transmitter and receiver and $c$ 
    is the speed of the light constant.
    
Measurement dataset~\cite{2jhd-wp15-19}  used in this study is obtained in an anechoic chamber that only allows \ac{los} propagation. Even though this setting moves the measurements away from rich scattering scenarios, the loss factors become a factor that can be defined as misalignment between antennas, hardware impairments, and path loss. Therefore, the received signal representation can be simplified to a combination of distance dependent path loss and misalignment between antennas. The effect of these losses on channel amplitude $a_{0}$ can be expressed as
\begin{equation}
    P_{rx}=P_{tx}+10  n \log _{10}(d)+ M,
    \label{path_loss}
\end{equation}
where $P_{rx}$ is the received power  calculated as the change in the power of the transmitted signal $P_{tx}$ due to path loss. Path loss exponent is denoted as $n$ and $M$ is the random antenna gain due to the effect of misalignment between antennas. 
Eq. \eqref{path_loss} expresses the received signal power in the LOS condition at the receiver that we are trying to model.
Since the signals under consideration are wideband and the channel is varied at \ac{thz} frequencies, $P_{rx}$ also fluctuates with frequency.
For this reason, in the next step, the \ac{gmm} is explained to better capture and model the power clusters and differences in $P_{rx}$.
\subsection{Gamma Mixture Model}
Let us define  two parameter $\alpha >0$ and $\beta>0$. The Gamma function $\Gamma(\alpha)$ is defined as
\begin{equation}
    \Gamma(\alpha)=\int_{0}^{\infty} x^{\alpha-1} e^{-x} d x.
\label{eq:Gamma_func}
\end{equation}
If both sides of Eq. \eqref{eq:Gamma_func} are divided by $\Gamma(\alpha)$ and by changing of variables as $x=\beta y$ follows

\begin{equation}
1=\int_{0}^{\infty} \frac{1}{\Gamma(\alpha)} x^{\alpha-1} e^{-x} d x=\int_{0}^{\infty} \frac{\beta^{\alpha}}{\Gamma(\alpha)} y^{\alpha-1} e^{-\beta y} d y .
\end{equation}
Then probability density function $f(x| \alpha, \beta)$ of Gamma distribution can be defined as
\begin{equation}
f(x \mid \alpha, \beta)= \frac{\beta^{\alpha}}{\Gamma(\alpha)} x^{\alpha-1} e^{-\beta x}, \  x \geq 0, \ \alpha > 0, \ \beta > 0, 
\label{Gamma_pdf}
\end{equation}
where parameters shape $\alpha$ and rate $\beta$ for all positive values of $x$ and it sums to one. 

The finite \ac{gmm} with $K$ components can be written as
\begin{equation}
p\left( \mathbf{x} | \alpha_{1}, \beta_{1}, \pi_{1}, \dots,\alpha_{K}, \beta_{K},\pi_{K}\right)=\sum_{k=1}^{K} \pi_{k} \mathcal{G}\left(\alpha_{k}, \beta_{k}\right) ,
\end{equation}
where $\mathbf{x}=\left\{x_{1}, \ldots, x_{n}\right\}$ is the positive vector of observations, $\pi_k$ mixing proportions or weights of the $k^{th}$ mixture component that sum to one $\sum_{k=1}^{K} \pi_{k}=1$ and $\mathcal{G}$ denotes Gamma distribution defined in Eq. \eqref{Gamma_pdf} with parameters $\alpha_k$ and $\beta_k$ which are the shape and rate parameters of the $k^{th}$ mixture component, respectively. 
The Gamma mixture was chosen for this study because it has traceable \ac{cdf} and MGF, thus it can provide an approximation for small-scale fading channels~\cite{atapattu2011mixture}.
Furthermore, by adjusting its parameters, a wide range of distributions can be represented with high accuracy and as showed in \cite{wiper2001mixtures}, arbitrary PDFs on $(0, \infty)$ can be modeled using Gamma mixture.
As a result, $P_{rx} = \{x_1, \dots, x_n\}$ can be modeled as a Gamma mixture, so we will need to find the appropriate number of Gamma clusters and their weights in addition to the parameters associated with each Gamma cluster. For ease of expression, the Gamma distribution parameters will be expressed together as $\theta_k= \{\alpha_k, \beta_k\}$ and throughout the remainder of this study, methods are described to find the Gamma mixture parameters in a way that models the empirical distribution of $P_{rx}$ observation vector.

\subsection{Expectation Maximization}
When some data is absent or latent variables are present, an iterative process called \ac{em} which is commonly used for density estimation such as clustering for mixture models is used to obtain the maximum likelihood estimates of the parameters. In this context, \ac{em} algorithm is herein utilized to fit the empirical \ac{pdfs} and find Gamma mixture parameters of the $P_{rx}$ measurements.

When the implementation of the \ac{em} algorithm is considered, it is observed that the number of Gamma components to be used for modeling $P_{rx}$ measurements is a priori information for its two-step process; expectation (E-step) and maximization (M-step)~\cite{mclachlan2007algorithm}. 
\ac{em} iterates until it reaches the desired convergence point, which may not be optimal~ \cite{ng2012algorithm}, updating the randomly assigned values for the parameters of the mixture model $\theta_{1: M}=\left(\theta_{1}, \ldots, \theta_{M}\right)$  at each iteration.
The membership coefficients for the entire measurement set $(i=1, \ldots, N)$ and all mixture components $(k=1, \ldots, M)$ are computed in the E-step using the current parameters $\theta_{1: M}$ as follows
\begin{equation}
\phi_{i k}=\frac{\pi_{k} p_{k}\left(x_{i} \mid \theta_{k}\right)}{\sum_{k=1}^{M} \pi_{k} p_{k}\left(x_{i} \mid \theta_{k}\right)},
\end{equation}
where $\sum_{k=1}^{M} \phi_{i k}=1$. 
Then, using the measurement data and membership coefficients derived in the E-step, new parameter values $\alpha$, $\beta$, and $\pi$  in the mixture model can be obtained for each Gamma component by updating the equations in the M-step as follows \cite{vegas2014gamma}
\begin{equation}
\pi_{k}^{n e w}=\frac{\sum_{i=1}^{N} \phi_{i k}}{N},
\label{new_pi}
\end{equation}

\begin{equation}
\mathbb{E}\left[X_{k}\right]^{\text {new }}=\frac{\sum_{i=1}^{N} \phi_{i k} x_{i}}{\sum_{i=1}^{N} \phi_{i k}}=\alpha \beta,
\label{a_b}
\end{equation}

\begin{equation}
\operatorname{Var}\left[X_{k}\right]^{n e w}=\frac{\sum_{i=1}^{N} \phi_{i k}\left(x_{i}-\mathbb{E}\left[X_{k}\right]^{n e w}\right)^{2}}{\sum_{i=1}^{N} \phi_{i k}}=\alpha \beta^{2} .
\label{a_b_2}
\end{equation}
Using the mixture parameters identified by Eq. \eqref{new_pi}, Eq. \eqref{a_b} and Eq. \eqref{a_b_2} we can represent the $P_{rx}$ observation vector in terms of a mixture of Gamma distribution.
\section{Dirichlet Process Gamma Mixture Model}
\label{sec:dpgmm}
\subsection{Dirichlet Process}
Dirichlet distribution is convenient in describing random probability mass functions for finite categorical sets, and it can be thought as a generalization of Beta distribution for multivariate data sets. 
For $K$ categorical event, Dirichlet distribution denoted as  $Dir(a_1, \ldots, a_K)$ can be given as
\begin{equation}
f\left(x_{1}, \ldots, x_{K} ; a_{1}, \ldots, a_{K}\right)=\frac{\Gamma\left(\sum_{i=1}^{K} a_{i}\right)}{\prod_{i=1}^{K} \Gamma\left(a_{i}\right)} \prod_{i=1}^{K} x_{i}^{a_{i}-1}.
\end{equation}
where $x_i$ represents a specific category, $\sum_{i=1}^{K} x_{i}=1$ and $x_{i} \geq 0$ for all $i \in\{1, \ldots, K\}$ also $a_i$ denotes the intensity of the specified category.

The infinite dimensional extension of Dirichlet distribution is a stochastic process called as \ac{dp}, whose realizations are probability distributions over some measurable set S.
Therefore, each draws from the \ac{dp}, itself a distribution.
This process is denoted as $DP(a, H)$, where $H$ is the expected value of the process called as base distribution and where $a$ is positive real number again describing the intensity of mass around the mean called as concentration parameter. 
If $G$ is a random variable drawn from $DP(a, H)$, then it can be shown that $G\left(A_{1}\right), \ldots, G \left(A_{n}\right) \sim {Dir}\left(a H\left(A_{1}\right), \ldots, a H\left(A_{n}\right)\right)$, where $\left\{A_{i}\right\}_{i=1}^{n}$ denotes any measurable finite partition of measurable set $S$ \cite{ferguson1973bayesian}.

The distinctive characteristic of the \ac{dp} is that the required number of clusters is obtained from the process due to nonparametric nature which makes it an ideal candidate for clustering problems where the number of clusters in unknown a priori.
The \ac{dp} is commonly used in non-parametric Bayesian models as a prior on the distribution space, especially in \ac{dpmm} which is known as infinite mixture models.
In our analysis, we can use \ac{dp} as a prior in distribution space to obtain the underlying distributions of the $P_{rx}$ values and their corresponding parameters.


\subsection{Stick-Breaking Construction of Dirichlet Process}

\ac{dp} can be built in a variety of methods such as the Blackwell–MacQueen urn scheme, the Chinese restaurant process, or the stick-breaking construction, and each of them emphasizes a distinct aspect of the \ac{dp}. 
In this study, we employed the stick-breaking construction, which allows expressing a sample from a \ac{dp} ; $G \sim DP(a,H)$ as \cite{sethuraman1994constructive}

\begin{equation}
G=\sum_{k=1}^{\infty} \pi_{k} \delta_{\theta_{k}}(\cdot), \quad \theta_{k} \sim H ,
\label{G}
\end{equation}
where $\theta_k$ represents atoms drawn independently from the base distribution $H$, $\delta_{\theta}$ point mass at $\theta$ and $\pi_k$ is the probability mass at atom $\theta_k$. The probability masses also known as weights and can be constructed as follows
\begin{equation}
    \pi_{k}=V_{k} \prod_{j=1}^{k-1}\left(1-V_{j}\right), \quad V_{k} \sim \mathcal{B}(1, a),
\end{equation}
 where $V_i$ denotes a broken piece, $a$ is the concentration parameter and $\mathcal{B}$ denotes Beta distribution.

This entire process may be illustrated by first breaking a unit-length stick randomly drawn by Beta distribution as, $V_1$ and then continuing to break the remaining portion of the stick $1-V_i$ randomly drawn by Beta distribution as $V_2 \dots V_k$. 
The weights are indicated by the length of each broken piece as $\pi_1 = V_1, \pi_2 = V_2 (1-V_1), \dots, \pi_k = V_k (1-V_{k-1})$.
Then atoms $\theta_k$ are drawn from the base distribution $H$ to associate these weights.
Upon breaking the stick, it becomes shorter, and the length of the higher indexed atoms decreases stochastically, whose rate of decrease depends on the \ac{dp} concentration parameter $a$ because it plays a key role in determining the weights in each iteration.
This procedure assures that $\sum_{k=1}^{\infty} \pi_{k}=1$. 
Note that for $G$ to be true from a $DP$, an infinite number of weights and atoms must be drawn.
However, in practice, it is possible to truncate summation on Eq. \eqref{G} with only a finite number of $K$ draws while still providing a very good approximation.
In Section \ref{sec:results}, we will go through how to choose the truncation number.
\subsection{Dirichlet Process Gamma Mixture Model}
Stick breaking construction of the \ac{dp} shows that samples from the process are discrete distribution. 
Thus, actually, \ac{dp} is not a proper prior for continuous distributions. 
Therefore, in non-parametric density estimation \ac{dp} is used indirectly by supporting kernel function $\mathcal{K}(\cdot)$ as described by
\begin{equation}
f(\mathbf{x}) =\int \mathcal{K}\left(x \mid \theta\right) G(\theta) \mathrm{d} \theta. \\
\label{eq:infinite_mixture_kernel}
\end{equation}
Choosing a \ac{dp} prior on $G$ as Eq. \eqref{G} and using it in Eq. \eqref{eq:infinite_mixture_kernel} can be turned into a sum of the infinite mixture of kernels \cite{gelman2013bayesian} 

\begin{equation}
f(\mathbf{x})=\sum_{k=1}^{\infty} \pi_{k} \mathcal{K}\left( x \mid \theta_{k}\right),
\end{equation}
which can be denoted as \ac{dpmm}. 
In Eq. \eqref{eq:infinite_mixture_kernel} the summation is set to infinity. This allows the model to describe new mixture components that may occur as new classes are added corresponding to the sampled atoms. 
However, it doesn't imply that infinitely many components are occupied. 

 Instead of simply sampling the data from the \ac{dp}, sampling the mixture parameters $\theta_i$ from the \ac{dp} and then utilizing these values as input in the continuous kernel function $\mathcal{K}(\cdot)$ enables the construction of \ac{dpmm} for non-parametric density estimation as hierarchically defined below
\begin{equation}
\begin{aligned}
x_{i} & \sim \mathcal{K}\left(\theta_{k}\right), \\
\theta_{k} & \sim G, \\
G & \sim \mathrm{DP}\left(a, H\right) . 
\end{aligned}
\end{equation}

 Thus, \ac{dpgmm}  can be defined by setting the kernel function $\mathcal{K}(\cdot)$ as  Gamma distributed
 \begin{equation}
     \mathcal{K}( x \mid \theta_k)  \equiv \mathcal{G}(x \mid \alpha_k, \beta_k),
 \end{equation}
 then it gets into the form of
 \begin{equation}
f(\mathbf{x})=\sum_{k=1}^{\infty} \pi_{k} \mathcal{G}\left(x \mid \alpha_{k}, \beta_{k}\right). 
\end{equation}
In order to find Gamma mixture parameters $\theta_k$,  base distribution $H$ and its hyperparameters needs to be defined for \ac{dpgmm}. 
To that end a modified hierarchical model out of \cite{elguebaly2015hierarchical} is utilized to estimate the mixture component parameters with fundamental modifications in terms of construction of \ac{dp}, defining priors and hyperpriors, and computing algorithm.

Assuming that $\alpha$ and $\beta$ follow prior distributions; $p(\alpha)$ and $p(\beta)$ respectively depending on hyperparameters; $\lambda, \kappa, \nu, v$. 
In order to add more flexibility to the model, we assume that also these hyperparameters follow some distributions; $p(\lambda), p(\kappa), p(\nu)$ and $p(v)$ depending on hyper-priors; $\vartheta, \varpi ,\varsigma, \varepsilon, \mu, \vartheta, \varrho$. 

The prior distributions for the mixture component parameters $\alpha$ and $\beta$ are all assumed independent of each other.
The Inverse-Gamma distribution ($\mathcal{IG}$) is used as a prior for the shape parameter $\alpha$, and hyperparameters with shape parameter $\lambda$ and ratio parameter $\kappa$ are utilized for this prior
\begin{equation}
p\left(\alpha_{k} \mid  \lambda, \kappa \right) \sim \frac{\kappa^{\lambda} \alpha_{k}^{-\lambda-1} \mathrm{e}^{-\kappa / \alpha_{k}}}{\Gamma(\lambda)}.
\end{equation}

An $\mathcal{IG}$ prior with hyperpriors $\vartheta$ and $\varpi$ are employed for $\lambda$ and an exponential ($\mathcal{E} x p$) prior with hyperprior $\varsigma$ is used for $\kappa$
\begin{equation}
p(\lambda \mid \vartheta, \varpi) \sim \frac{\varpi^{\vartheta} \mathrm{e}^{-\varpi / \lambda}}{\Gamma(\vartheta) \lambda^{\vartheta+1}}  , 
\end{equation}
\begin{equation}
p(\kappa \mid \varsigma) \sim \zeta \mathrm{e}^{-\varsigma \kappa}.
\end{equation}
With these above specifications for the hyperparameter $\alpha$, its posterior can cover a very large range in positive axis with given vague hyperpriors on $\vartheta,\varpi$, and $\varsigma$. 

Also, Gamma distribution is used as a prior for the rate parameter $\beta$, and hyperparameters with shape parameter $\nu$ and ratio parameter $v$ are utilized for this prior
\begin{equation}
p\left(\beta_{k} \mid \nu, v  \right) \sim \frac{v^{\nu}}{\Gamma(\nu)} \beta_{k}^{\nu-1} \mathrm{e}^{-v \beta_{k}}.
\end{equation}

An $\mathcal{IG}$ prior with hyperpriors $\epsilon$ and $\mu$  is used for $v$ and a $\mathcal{G}$ prior with hyperpriors $\varphi$ and $\varrho$ is used for $\nu$
\begin{equation}
p(\nu \mid \epsilon, \mu) \sim \frac{\mu^{\epsilon} \mathrm{e}^{-\mu / \nu}}{\Gamma(\epsilon) \nu^{\epsilon+1}},
\end{equation}
\begin{equation}
p(v \mid \varphi, \varrho) \sim \frac{v^{\varphi-1} \varrho^{\varphi} \mathrm{e}^{-\varrho v}}{\Gamma(\varphi)}.
\end{equation}

Based on these specifications for the hyperparameter $\beta$, its posterior can cover especially small values between 0 and 1 with given vague hyperpriors on $\epsilon,\mu $ and $\varphi, \varrho$, which are specified in Section IV-B. 

In addition to base distribution also concentration parameter $a$ which also governs the distribution over the number of components $K$, has to be defined for the \ac{dpgmm}. In \cite{elguebaly2015hierarchical}, Inverse-Gamma distribution was chosen as a prior to $a$. However, utilization of  $\mathcal{IG}$ leads to the inclusion of an excessive number of mixture components in the estimation process thus estimation flexibility becomes limited. Thus, in this study, conjugacy of the Gamma distribution to the $\mathcal{B}(1, a)$ is exploited to provide flexibility to the cluster sizes and mixture weights. So, $\mathcal{G} (1, 1)$ is used as a prior for concentration parameter $a$, therefore it became possible to represent distributions using fewer components while maintaining estimation accuracy. Please note that the proposed model's hierarchical structure allows flexibility and adaptability to a wide range of data with various distributional characteristics.
Furthermore, finding robust parameters for models may be challenging, and misspecified parameters may diminish model performance; consequently, using a hierarchical model alleviates this difficulty while also providing flexibility and robustness to the model framework \cite{gorur2010Dirichlet}.

\begin{figure*}[!ht]
    \centering
    \includegraphics[width=17cm]{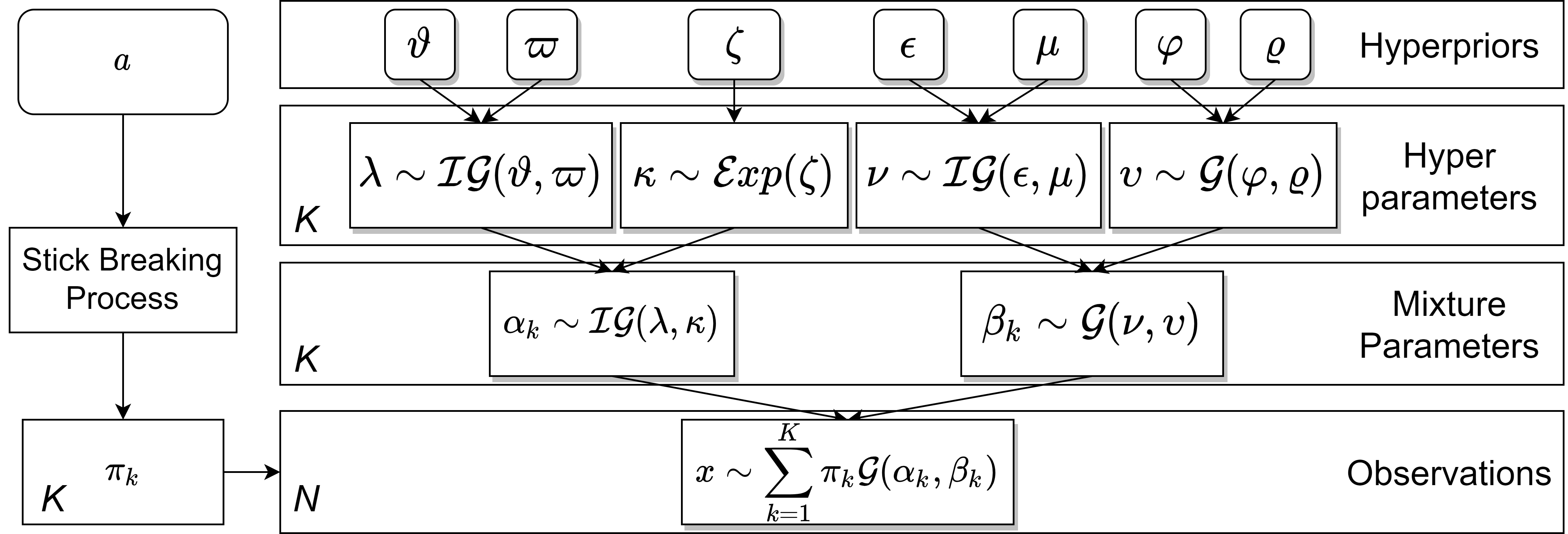}
    \caption{Hierarchical Dirichlet process Gamma mixture model structure.}
    \label{fig:hierarchical_structure}
\end{figure*}

The structure of our model summarized in Fig. \ref{fig:hierarchical_structure} and \ac{dpgmm} can be described all together hierarchically using the previously mentioned structure of the weights, base distribution, and the concentration parameter as described below:
\begin{subequations}
\begin{align}
a & \sim \mathcal{G}(1,1), \\
V_{1}, \ldots, V_{K} & \sim \mathcal{B}(1, a), \\
\pi_{k} &=V_{k} \prod_{j=1}^{k-1}\left(1-V_{j}\right),\\
\lambda_{k} &\sim \mathcal{IG}(\vartheta, \varpi),\\
\kappa_{k} &\sim \mathcal{E} x p(\varsigma), \\
\nu_{k} &\sim  \mathcal{G}(\epsilon, \mu), \\
v_{k} &\sim \mathcal{IG}( \varphi, \varrho), \\
\alpha_k &\sim \mathcal{IG}(\lambda_{k}, \kappa_{k} \mid  \vartheta, \varpi, \varsigma)  \\ 
\beta_k &\sim \mathcal{G} (\nu_{k}, v_{k} \mid \epsilon, \mu, \varphi, \varrho ), \\
x_i & \sim \sum_{k=1}^{K} \pi_{k} \mathcal{G}\left(x \mid \alpha_k, \beta_k \right).
\end{align}
\label{eq:hieararchical_model}
\end{subequations}

Since our prior assumptions for mixture parameters independent of each other, a draw from the base distribution $(\alpha_k, \beta_k) \sim H(\theta_k)$ by Eq. (28h) and Eq. (28i) can be defined as 
\begin{equation}
    \alpha_k, \beta_k \sim \mathcal{IG}(\lambda_{k}, \kappa_{k} \mid  \vartheta, \varpi, \varsigma) \mathcal{G} (\nu_{k}, v_{k} \mid \epsilon, \mu, \varphi, \varrho ).
\end{equation}
We aim to find posterior distribution for the \ac{dpgmm}.
A mixture model which gives the best evidence of assigning each data point $\mathbf{x} = ({x_1, \dots, x_i})$ to only one cluster, can be found by selecting the model that maximizes the integrated likelihood, which is  \cite{biernacki2000assessing}
\begin{equation}
\mathbf{f}(\mathbf{x}, \mathbf{z} \mid \psi)=\int_{\Theta} \mathbf{f}(\mathbf{x}, \mathbf{z} \mid \theta, \psi) p(\theta \mid \psi)d \theta,
\label{eq:integrated_likelihood_integral}
\end{equation}
with 
\begin{equation}
\mathbf{f}(\mathbf{x}, \mathbf{z} \mid \theta, \psi)=\prod_{i=1}^{n} f\left({x}_{i}, {z}_{i} \mid \theta, \psi \right),
\label{eq:integrated_likelihood_1}
\end{equation}
and 
\begin{equation}
f\left({x}_{i}, {z}_{i} \mid \theta, \psi\right)=\prod_{k=1}^{K} \left(\pi_{k} p\left(x_{i} \mid \theta_{k}, \psi_{k} \right)\right)^{\mathbb{I}\left(z_{i}=k\right)},
\label{eq:integrated_likelihood_2}
\end{equation}
where $\Theta=\{{\theta_{1:K}}, { \pi_{1:K}}, \psi_{1:K} \}$ denotes parameter space of the model, $\theta_k = \{ \alpha_k, \beta_k \}$ and $\psi_{k}= \{ \lambda_{k}, \kappa_{k}, \mu_{k}, \nu_{k}\}$ denotes hyperparameters. 
In addition, $z_i$ is an indicator variable that describes the assignment of the observation $x_i$ to a specific mixture component between $1$ and $K$. 
It is also known as a latent variable or missing allocation variable introduced by the nature of \ac{dp} clustering property and its incidence represents formation of a new component on the mixture, which is controlled by $a$. 
The indicator function for $z_i$ is denoted by $\mathbb{I}$. 
By following Eq. \eqref{eq:integrated_likelihood_1} and Eq. \eqref{eq:integrated_likelihood_2}, the likelihood function for the model can be given as

\begin{equation}
p(\mathbf{x}, \mathbf{z} \mid \Theta)=\prod_{i=1}^{n} \prod_{k=1}^{K}\left( \pi_k \  p\left(x_{i} \mid \theta_{k}, \psi_k\right) \right)^{\mathbb{I}\left(z_{i}=k\right)},
\label{eq:likelihood}
\end{equation} 
Hence by using the Bayes rule, the posterior probability is equal to the likelihood times the prior divided by the evidence
\begin{equation}
p(\Theta \mid \mathbf{x}, \mathbf{z})=\frac{p(\Theta) p(\mathbf{x}, \mathbf{z} \mid \Theta)}{\int p(\Theta) p(\mathbf{x}, \mathbf{z} \mid \Theta) d\Theta  } \propto p(\Theta) p(\mathbf{x}, \mathbf{z} \mid \Theta).
\label{eq:bayes}
\end{equation}

Also, posterior density proportional to joint probability, $p(\Theta \mid \mathbf{x}, \mathbf{z}) \propto p(\mathbf{x}, \mathbf{z}, \Theta)$ where the marginal likelihood in the denominator in Eq. \eqref{eq:bayes}, can be disregarded as it does not depend on the model parameters.
In a sense, determining the joint probability function provides information on the posterior density.
We can define the joint probability of the model taking into account hyperparameters and hyperpriors as~\cite{zobay2009mean}
\begin{equation}
\begin{aligned}
&p\left(\mathbf{x}, z_{1:K}, {\theta_{1:K}}, {\pi_{1:K}}, \psi_{1:K}\right) = \\
&\prod_{i=1}^{n} \prod_{k=1}^{K}  \left[ \pi_{k}  \ p\left(x_{i} \mid \theta_{k}, \psi_k\right) \right]^{\mathbb{I}\left(z_{i}=k\right)} \times\\
&\prod_{k=1}^{K} p\left(\alpha_{k}, \beta_k \mid \psi_k \right) \prod_{k=1}^{K-1} \mathcal{B}\left(1, a \right).
\label{eq:joint_probability}
\end{aligned}
\end{equation}

When dealing with \ac{dpgmm}s, the posterior distributions and posterior of the model parameters are usually analytically intractable when a non-conjugate base distribution is chosen for the mixture kernel due to Eq. \eqref{eq:bayes} involves integration. 
However, by having 
$p(\mathbf{x}, \mathbf{z}, \Theta)$ as
Eq. \eqref{eq:joint_probability} we can simulate our posterior distribution and model parameters $\Theta$, rather than computing them. 
Hence, the inference is done through computational \ac{mcmc} simulations. 
To generate samples from \ac{dpgmm} posterior distributions, several \ac{mcmc} simulation algorithms based on Gibbs sampling, Metropolis-Hastings and \ac{hmc} have been developed, which can subsequently be utilized for inference of all parameters and posterior distribution. 

Furthermore, the \ac{nuts} \cite{hoffman2014no} a variant of the \ac{hmc} that is a newer and more efficient algorithm than the others, can also be used for posterior sampling.
By leveraging first-order gradient information, \ac{nuts} and \ac{hmc} provide faster convergence than other sampling methods, especially in complex and high-dimensional datasets. 
However, unlike \ac{hmc}, \ac{nuts} contains many self-tuning procedures for adaptively adjusting the tunable parameters of \ac{hmc} like step size and desired number of steps. 
We also should mention that, the \ac{nuts} algorithm is available in the PyMC3 \cite{salvatier2016probabilistic}
which is a python based probabilistic programming package, and it allows fitting Bayesian models with a variety of \ac{mcmc} simulation algorithms. Moreover, it provides extra advantages in gradient computation.
For these particular and distinctive reasons, the PyMC3 package with the \ac{nuts} algorithm is applied to posterior sampling in this study.

\section{Gamma Mixture Model for Terahertz Band Wireless Channels}
\label{sec:simulations_and_results}

\begin{figure*}[!ht]
  \centering
  \subfigure[]{\includegraphics[width=5.86cm]{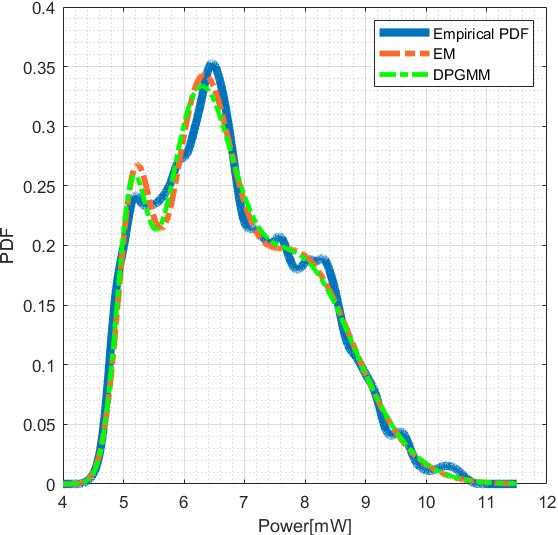}}
  \subfigure[]{\includegraphics[width=5.86cm]{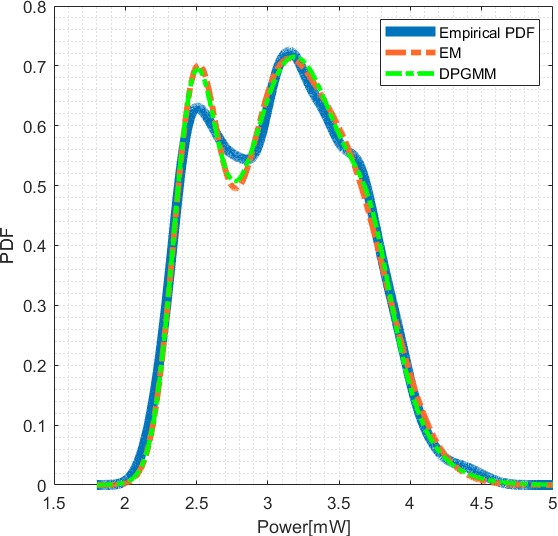}}
  \subfigure[]{\includegraphics[width=5.86cm]{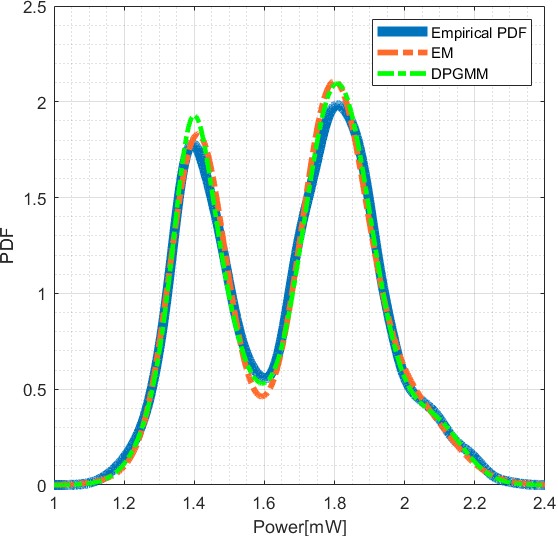}}
  
  \centering
  \subfigure[]{\includegraphics[width=5.86cm]{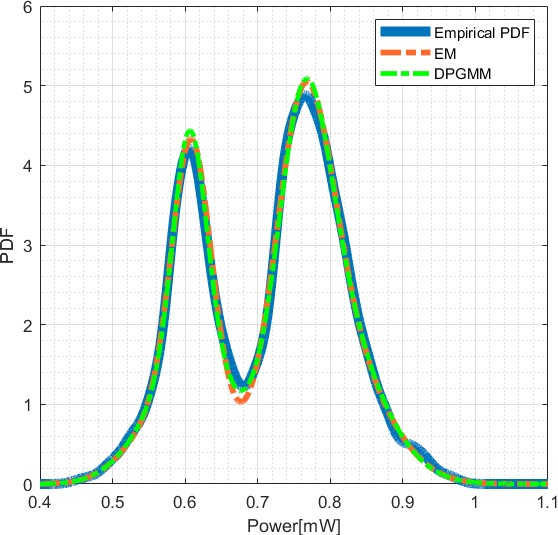}}
  \centering
  \subfigure[]{\includegraphics[width=5.86cm]{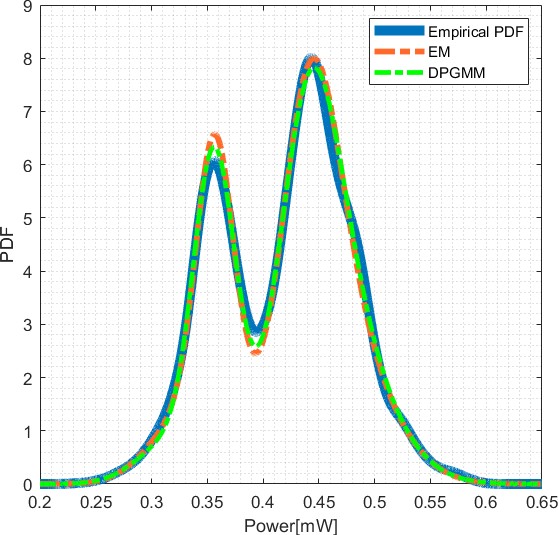}}
  \caption{DPGMM and \ac{em} Gamma mixture model for emprical \ac{pdfs}
  (a) 20 cm, (b) 30 cm, (c) 40 cm, (d) 60 cm, (e) 80 cm. }
  \label{fig:estimates}
\end{figure*}
In this section, first the measurement setup and the datasets will be briefly described in Section \ref{sec:measurement_dataset}. 
After explaining pre-processing of the dataset and post-processing settings of the \ac{dpgmm} and \ac{em} algorithm in Section \ref{sec:simulation_settings}, we are presented the error metric \ac{kl} in Section \ref{sec:error} to quantify performance with respect to the experimental \ac{pdfs} with the proposed approaches driven by the simulation settings.
Finally, in Section \ref{sec:results}, we will discuss and compare the results of the \ac{dpgmm} and \ac{em} algorithm.

\subsection{Measurement Dataset}
\label{sec:measurement_dataset}
The measurements utilized in this paper were obtained at the Turkish Science Foundation's (T\"UB\.ITAK) anechoic chamber \cite{Millimet77:online}, which are also available at \cite{2jhd-wp15-19}.
The data consist of complex $S_{21}$ parameters taken at five different points away from the transmitter \textit{i.e.}, at 20 cm, 30 cm, 40 cm, 60 cm, and 80 cm respectively. 
A laser-based system is used for measurements to precisely align the transmitter and receiver, ensuring \ac{los} condition and measurement reliability.
Measurements cover the 60 GHz band between 240-300 GHz with 4096 points of $S_{21}$ measurements. 
The spectrum resolution becomes $14.468$~MHz with this configuration.
Also, IF bandwidth is set to 100~Hz for the measurements, which allows for an increase in the observed dynamic range and a decrease in the noise floor.
Detailed information about measurement setup can be found in \cite{tekbiyik2021modeling}.

 One important thing to mention, due to the short wavelengths of THz frequencies, smaller antennas will be utilized, which will also result in an expansion of communication applications in next-generation communication systems. One such area that THz frequencies will enable is the application of microscale communications\cite{he2017stochastic}. Microscale communication applications include high-speed wireless connections between personal mobile terminals, PC/kiosk/cloud servers, and wireless nanosensor communications, which link tiny devices across considerably shorter distances \cite{akyildiz2010electromagnetic}.
 The measurement results we employed in this study are consistent with microscale THz communication scenarios.

\subsection{Data Processing and Simulation Settings}
\label{sec:simulation_settings}
\begin{table*}[!ht]
\centering
\renewcommand{\arraystretch}{1.4}
\setlength{\tabcolsep}{14pt}
\caption{Mixture parameters and error metrics for PDF estimations at distinct distances.}
\label{tab:parameters_error_metrics}
\begin{tabular}{|c|c|c|ccc|c|}
\hline
\multirow{2}{*}{\textbf{Distance}} & \multirow{2}{*}{\textit{\textbf{K}}} & \multirow{2}{*}{\textbf{Method}} & \multicolumn{3}{c|}{\textbf{Mixture Parameters}} & \multirow{2}{*}{\textbf{KL-Divergence}} \\ \cline{4-6}
                                  &                                      &                                  & $\pi$          & $\alpha$        & $\beta$       &                                         \\ \hline
\multirow{6}{*}{20 cm}             & \multirow{6}{*}{3}                   & \multirow{3}{*}{EM}              & 0.41942        & 141.51136       & 0.04461       & \multirow{3}{*}{0.03343}                \\
                                  &                                      &                                  & 0.41571        & 82.46476        & 0.09704       &                                         \\
                                  &                                      &                                  & 0.16485        & 345.43516       & 0.01502       &                                         \\ \cline{3-7} 
                                  &                                      & \multirow{3}{*}{DPGMM}           & 0.47049        & 108.87149       & 0.05768       & \multirow{3}{*}{0.03603}                \\
                                  &                                      &                                  & 0.39966        & 83.45134        & 0.09634       &                                         \\
                                  &                                      &                                  & 0.12972        & 424.98806       & 0.01208       &                                         \\ \hline
\multirow{6}{*}{30 cm}             & \multirow{6}{*}{3}                   & \multirow{3}{*}{EM}              & 0.39304        & 140.15848       & 0.02194       & \multirow{3}{*}{0.00581}                \\
                                  &                                      &                                  & 0.34110        & 145.96878       & 0.02469       &                                         \\
                                  &                                      &                                  & 0.26585        & 238.59034       & 0.01050       &                                         \\ \cline{3-7} 
                                  &                                      & \multirow{3}{*}{DPGMM}           & 0.72306        & 63.55631        & 0.05088       & \multirow{3}{*}{0.00378}                \\
                                  &                                      &                                  & 0.21586        & 270.00446       & 0.00921       &                                         \\
                                  &                                      &                                  & 0.06098        & 657.81747       & 0.00568       &                                         \\ \hline
\multirow{8}{*}{40 cm}             & \multirow{8}{*}{4}                   & \multirow{4}{*}{EM}              & 0.49563        & 353.30886       & 0.00510       & \multirow{4}{*}{0.00928}                \\
                                  &                                      &                                  & 0.28849        & 336.61776       & 0.00419       &                                         \\
                                  &                                      &                                  & 0.11952        & 105.15786       & 0.01376       &                                         \\
                                  &                                      &                                  & 0.09634        & 370.67005       & 0.00549       &                                         \\ \cline{3-7} 
                                  &                                      & \multirow{4}{*}{DPGMM}           & 0.52684        & 323.77472       & 0.00560       & \multirow{4}{*}{0.00860}                \\
                                  &                                      &                                  & 0.28225        & 159.00929       & 0.00901       &                                         \\
                                  &                                      &                                  & 0.12817        & 894.49836       & 0.00155       &                                         \\
                                  &                                      &                                  & 0.06264        & 658.57817       & 0.00317       &                                         \\ \hline
\multirow{8}{*}{60 cm}             & \multirow{8}{*}{4}                   & \multirow{4}{*}{EM}              & 0.43426        & 366.50425       & 0.00208       & \multirow{4}{*}{0.07901}                \\
                                  &                                      &                                  & 0.21705        & 522.76836       & 0.00116       &                                         \\
                                  &                                      &                                  & 0.17665        & 218.71135       & 0.00380       &                                         \\
                                  &                                      &                                  & 0.17202        & 93.00082        & 0.00648       &                                         \\ \cline{3-7} 
                                  &                                      & \multirow{4}{*}{DPGMM}           & 0.31911        & 480.53104       & 0.00159       & \multirow{4}{*}{0.07699}                \\
                                  &                                      &                                  & 0.27953        & 194.58752       & 0.00419       &                                         \\
                                  &                                      &                                  & 0.21470        & 91.27561        & 0.00668       &                                         \\
                                  &                                      &                                  & 0.18655        & 671.84839       & 0.00090       &                                         \\ \hline
\multirow{8}{*}{80 cm}             & \multirow{8}{*}{4}                   & \multirow{4}{*}{EM}              & 0.42664        & 287.17883       & 0.00154       & \multirow{4}{*}{0.12921}                \\
                                  &                                      &                                  & 0.21421        & 454.25301       & 0.00078       &                                         \\
                                  &                                      &                                  & 0.20415        & 162.81782       & 0.00295       &                                         \\
                                  &                                      &                                  & 0.15497        & 73.67085        & 0.00474       &                                         \\ \cline{3-7} 
                                  &                                      & \multirow{4}{*}{DPGMM}           & 0.32959        & 130.13821       & 0.00358       & \multirow{4}{*}{0.12347}                \\
                                  &                                      &                                  & 0.32336        & 287.00556       & 0.00154       &                                         \\
                                  &                                      &                                  & 0.25716        & 368.78112       & 0.00096       &                                         \\
                                  &                                      &                                  & 0.08985        & 79.30695        & 0.00421       &                                         \\ \hline
\end{tabular}
\end{table*}
First, by using $S_{21}$ parameters obtained from measurements, the received power $P_{rx}$ is calculated by
\begin{equation}
P_{rx}=\left|S_{21}\right|^{2} P_{tx},
\end{equation}
where $P_{tx}$ power of the transmitted signal and $\left|S_{21}\right|$ is amplitude response of transmission channel.
Second, the proposed models are used to estimate the underlying distributions of empirical measurement \ac{pdfs}, which are constructed from the histogram of $P_{rx}$ values.
In \ac{em}, mixture component number must be given a priori.
However, because the required number of mixture components is unknown beforehand, we used the number of mixture components obtained from \ac{dpgmm} in \ac{em} as input to perform the \ac{em} algorithm.
Note that also, instead of starting from random points to run the EM algorithm, the first estimate is obtained via k-means algorithm, which is then used by the EM algorithm to run its iterations \cite{vegas2014gamma}.
Then \ac{em} Gamma mixture parameters can be found by using Eqs. (11), (12), and (13) and with \ac{dpgmm}, parameters can be inferred using the hierarchical structure of Eqs. (28) by sampling from posterior by \ac{nuts}. 
Also, vague values given to the hyperpriors for \ac{dpgmm} as  $\vartheta = 1, \varpi = 1, \varsigma = 0.001,\epsilon = 1, \mu = 1,\varphi = 1, \varrho = 1 $. 

\subsection{Error Metric}
\label{sec:error}
In this paper, \ac{kl} is performed to compare the experimental distributions with predicted distributions using  \ac{dpgmm} and \ac{em}. The \ac{kl} is a metric that is used to assess the difference between two probability distributions over the same probability space ${\mathcal{X}}$. 
Let $Q(x)$ be the distribution whose distance from the reference distribution $P(x)$ is to be measured, then \ac{kl} can be given as
\begin{equation}
    \mathcal{D}_{K L}(P \| Q)=\sum_{x \in \mathcal{X}} P(x) \log \left(\frac{P(x)}{Q(x)}\right),
\end{equation}
where $P_x$ represent experimental distribution and $Q_x$ represent predicted distributions via \ac{em} algorithm and \ac{dpgmm}.

\subsection{Results and Comparison}
\label{sec:results}

The results of the \ac{em} algorithm and \ac{dpgmm} obtained by using the $P_{rx}$ histograms are presented in this section. In order to plot measurements and compare results, we transformed $P_{rx}$ histograms to the empirical \ac{pdfs}.

In Fig. \ref{fig:estimates} we showed estimated \ac{pdfs} utilizing \ac{dpgmm} and \ac{em} algorithm to the $P_{rx}$ values for five distinct distances. 
Different histograms are used to illustrate the flexibility and accuracy of the proposed model with a variety of distributions with different forms.
As can be seen from Fig. \ref{fig:estimates}, the \ac{dpgmm} and \ac{em} algorithm can describe all empirical \ac{pdfs} very well.
Although the hyperprior values are not reliant on the measurement data, they do incorporate the distributions acquired from the sub-THz band observations.
It can also cover a large distribution space specified on the positive axis.

Table \ref{tab:parameters_error_metrics} contains estimated parameter values of mixture components and \ac{kl} values for the \ac{dpgmm} and the \ac{em} algorithm. 
As shown in Table \ref{tab:parameters_error_metrics}, although there is a difference in the \ac{kl} values for \ac{dpgmm} and \ac{em} algorithm, it is not significant because the values are small and very close to each other. 
Furthermore, as seen in Fig. 1, we may neglect these discrepancies because the \ac{dpgmm} and \ac{em} algorithm both match the empirical \ac{pdfs} fairly well.
The important thing to be considered here is the ability of the \ac{dpgmm} to fit the empirical \ac{pdfs} as much as the \ac{em} algorithm, even if the required number of mixture components is not known a priori.

\begin{figure*}[!ht]
    \centering
    \includegraphics[width=15cm]{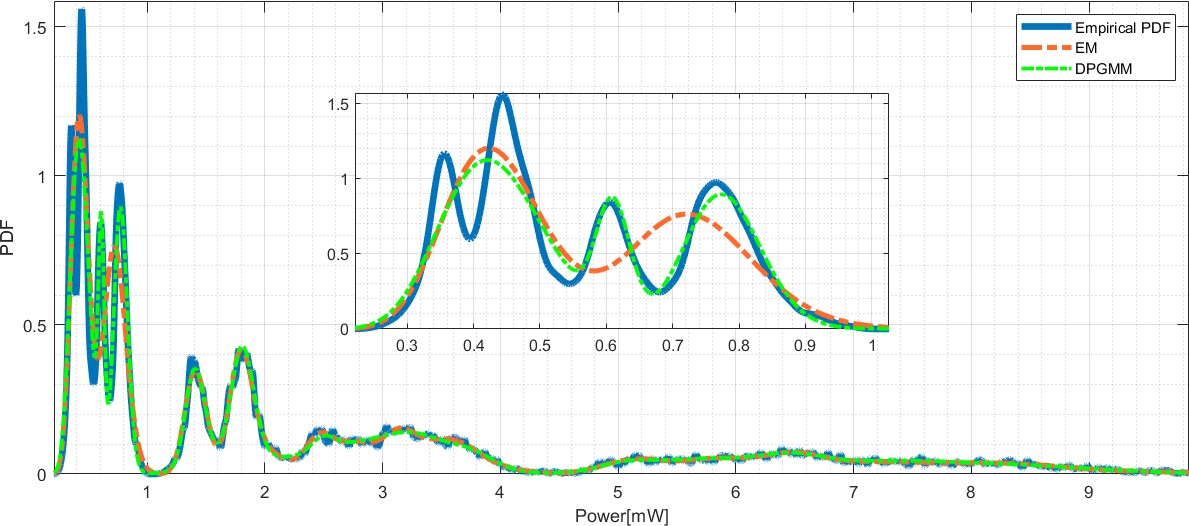}
    \caption{Empirical PDF for combination of all measurement data and Gamma mixture models.}
    \label{fig:all_measurement}
\end{figure*}

Furthermore, we combined the $P_{rx}$ values of all measurements collected at different distances and modeled the obtained PDF using the proposed methods as shown in Fig. \ref{fig:all_measurement}. 
In our case, the performance of the \ac{em} algorithm decreases when the data size and data dimensionality increase as given in Table \ref{tab:my-table}. 
\ac{em} algorithm was not able to model $0.5-0.7$ mW $P_{rx}$ region accurately, as can be seen in Fig. \ref{fig:all_measurement}.
One reason is that the \ac{em} algorithm converges to the local maxima rather than the global maxima. 
Also, the \ac{em} algorithm does not always guarantee convergence to the local maxima, it only guarantees convergence to a point with zero gradients according to the parameters and depending on the initialization step~\cite{karlis2003choosing}. 
As a result, the \ac{em} algorithm might occasionally become stuck at the saddle points \cite{ng2012algorithm,mclachlan2007algorithm}.
Besides, \ac{dpgmm} is able to model the combination of all $P_{rx}$ values very well.
\begin{table}[!ht]
\centering
\renewcommand{\arraystretch}{1.5}
\setlength{\tabcolsep}{10pt}
\caption{Error metrics for all measurement data}
\label{tab:my-table}
\begin{tabular}{|c|c|c|c|}
\hline
\textbf{Distance}    & \textit{\textbf{K}} & \textbf{Method} & \textbf{KL-Divergence} \\ \hline
\multirow{2}{*}{All} & \multirow{2}{*}{17} & EM              & 0.07697                \\ \cline{3-4} 
                     &                     & DPGMM           & 0.03759                \\ \hline
\end{tabular}
\end{table}

\begin{figure}[!ht]
    \centering
    \includegraphics[width=8.6cm]{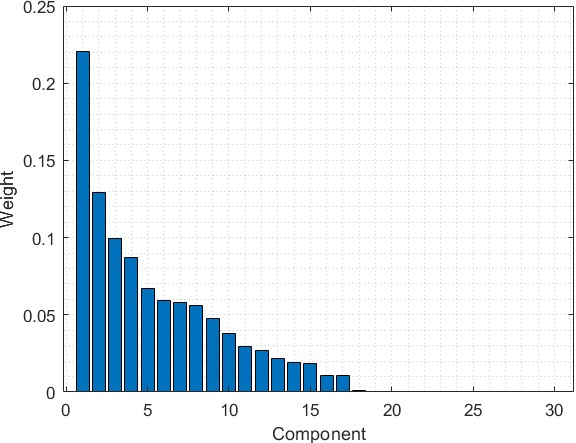}
    \caption{Number of required components inferred with \ac{dpgmm} for combination of all measurement data empirical PDF.}
    \label{fig:max_component}
\end{figure}

 In addition to these, we must address the truncation number $K$. 
 Normally \ac{dp} consists of an infinite number of distributions however, for computational convenience, $K$ should be chosen in a way that can accurately describe the \ac{dp}. 
 It has to be large enough to represent the empirical PDF of all measurement data while being cost-effective for posterior sampling. 
 For our \ac{dpgmm}, we choose $K=30$, which is sufficient for all measurement histograms. 
 This does not imply that the clusters occupy $K$ components in the data samples; instead, the model allows flexibility by introducing new mixture components up to 30, if necessary, as samples are added. 
 As an illustration, for the empirical PDF of all $P_{rx}$ data in Fig. \ref{fig:all_measurement}, it is sufficient to use a maximum of 17 mixture components with \ac{dpgmm} as shown in Fig. \ref{fig:max_component}.

Finally, one additional mixture component is added to the number of mixture components extracted using \ac{dpgmm} for each distance, and the \ac{em} algorithm is applied to the $P_{rx}$ measurements with this mixture component number. 
The \ac{kl} values for \ac{em} obtained with these settings are given in Table \ref{tab:extra_component_kl}, along with the \ac{kl} values of \ac{dpgmm}  obtained previously. 
Although the number of components increased by one, the \ac{kl} values of the \ac{em} algorithm vary very little, as shown in Table \ref{tab:extra_component_kl}.

This demonstrates that the number of mixture components provided by \ac{dpgmm} is sufficient for modeling empirical \ac{pdfs}, and adding more components will not significantly improve overall performance.

\begin{table}[!ht]
\centering
\renewcommand{\arraystretch}{1.5}
\setlength{\tabcolsep}{10pt}
\caption{The comparison of error metrics of the \ac{dpgmm} and \ac{em} algorithm which is initialized with one additional number of mixture component.}
\label{tab:extra_component_kl}
\begin{tabular}{|c|c|c|c|}
\hline
\textbf{Distance}      & \textit{\textbf{K}} & \textbf{Method} & \textbf{KL-Divergence} \\ \hline
\multirow{2}{*}{20 cm} & 4                   & EM              & 0.03298                \\ \cline{2-4} 
                       & 3                   & DPGMM           & 0.03603                \\ \hline
\multirow{2}{*}{30 cm} & 4                   & EM              & 0.00402                \\ \cline{2-4} 
                       & 3                   & DPGMM           & 0.00581                \\ \hline
\multirow{2}{*}{40 cm} & 5                   & EM              & 0.00937                \\ \cline{2-4} 
                       & 4                   & DPGMM           & 0.00860                \\ \hline
\multirow{2}{*}{60 cm} & 5                   & EM              & 0.07843                \\ \cline{2-4} 
                       & 4                   & DPGMM           & 0.07699                \\ \hline
\multirow{2}{*}{80 cm} & 5                   & EM              & 0.12915                \\ \cline{2-4} 
                       & 4                   & DPGMM           & 0.12347                \\ \hline
\end{tabular}
\end{table}

\subsection{Processing Times}


Since \ac{dpgmm} uses a \ac{mcmc} based sampling algorithm, as a result, we use this algorithm for a situation where we have no prior knowledge, so we start solving the problem by considering certain points as the starting point. This initial state may be close or far from the result, so we move posterior by taking each sample and updating our estimates according to the Bayesian probability with the observations.  
The processing time of \ac{mcmc} algorithms increases as the model parameters increase, this is firstly because of the dimensionality, where the volume of the sample space increases with the number of parameters \cite{brooks1998markov}. As a result, the \ac{mcmc} algorithm is computationally heavier than the EM algorithm \cite{ryden2008versus}. The processing times are given in the Table \ref{tab:processing_time}. Note that, the dataset contains all measurements consist of 10240 data points others are 2048 points.

\begin{table}[ht!]
\centering
\renewcommand{\arraystretch}{1.5}
\setlength{\tabcolsep}{2pt}
\caption{Processing Times Of EM Algorithm and DPGMM (with MCMC) for different dataset}
\label{tab:processing_time}
\begin{tabular}{|l|c|c|c|c|c|c|}
\hline
               & \textbf{20 cm} & \textbf{30 cm} & \textbf{40 cm} & \textbf{60 cm} & \textbf{80 cm} & \textbf{All} \\ \hline
\textbf{EM}    & 8.98s          & 9.45s          & 9.31s          & 6.60s          & 6.13s          & 22.97s       \\ \hline
\textbf{DPGMM} & 32m25.4s      & 25m15.4s       & 21m29.4s       & 9m13.4s        & 22m25.4s       & 179m53.4s    \\ \hline
\end{tabular}
\end{table}

However, it should also be taken into account that, in contrast to \ac{mcmc}, for the \ac{em} algorithm we give the starting points by using the k-means algorithm instead of giving the starting points randomly so that it can converge to the target distribution faster with the number of mixture components that should be given as a priori in the EM algorithm. This allows the \ac{em} algorithm to run faster and guarantees to convergence the result. Otherwise, the EM algorithm cannot converge, or the density estimation performance is insufficient. 

To minimize calculation time, the variational inference approach can be applied to the same model.
Variational inference delivers solutions with equivalent accuracy to \ac{mcmc} sampling at a faster rate in many situations\cite{blei2017variational}.

\section{ Concluding Remarks and Future Works}

In this work, we proposed utilizing a flexible hierarchical \ac{dpgmm} for the sub-THz wide-band channel modeling. 
The main reasons that Gamma distribution is chosen as a kernel for the mixture are because of its flexibility, skewness, ability the model tailed distributions and moreover its \ac{cdf}, MGF and moments are tractable.
The proposed model is applied to the measurement dataset, which is between 240 GHz and 300 GHz.  
Also \ac{em} algorithm is utilized for the same dataset in order to compare the parameter estimation performance of \ac{dpgmm} and \ac{em} algorithm. 
In the \ac{em} algorithm the required number of mixture components must be known a priori, however the proposed \ac{dpgmm} is able to infer the required number of mixture components and corresponding component parameters.   
The simulation results reveal that the proposed \ac{dpgmm} can accurately describe the various sub-THz channels that occur as a result of different scenarios, as much as the \ac{em} algorithm. 
However, for the higher dimensional histogram, \ac{dpgmm} is able to describe underneath distributions better than the \ac{em} algorithm.  
Due to the given hyperprior specifications and hierarchical structure of the model, it is shown that \ac{dpgmm} can be used for any type of wireless channel, not just for the \ac{thz} band.

For Gamma mixture wireless channels, average channel capacity, the outage probability, and the symbol error rate were derived earlier. 
By making use of this provided knowledge, analytical analyzes can be made as future work using the mixture parameters given in this study. 
In addition to future works, measurements taken in different scenarios such as misalignment fading can also be modeled with given \ac{dpgmm}.
We believe that this work paves the way for modeling and performance analysis of not only the \ac{thz} band also any kind of wireless communication channel in 5G and 6G networks.
\section*{Acknowledgments}
The authors would like to thank Ali Rıza Ekti and K\"ur\c{s}at Tekb{\i}y{\i}k for their assistance.

\bibliographystyle{IEEEtran}
\bibliography{main.bib}


\begin{IEEEbiography}[{\includegraphics[width=1.1in,height=1.3in,clip,keepaspectratio]{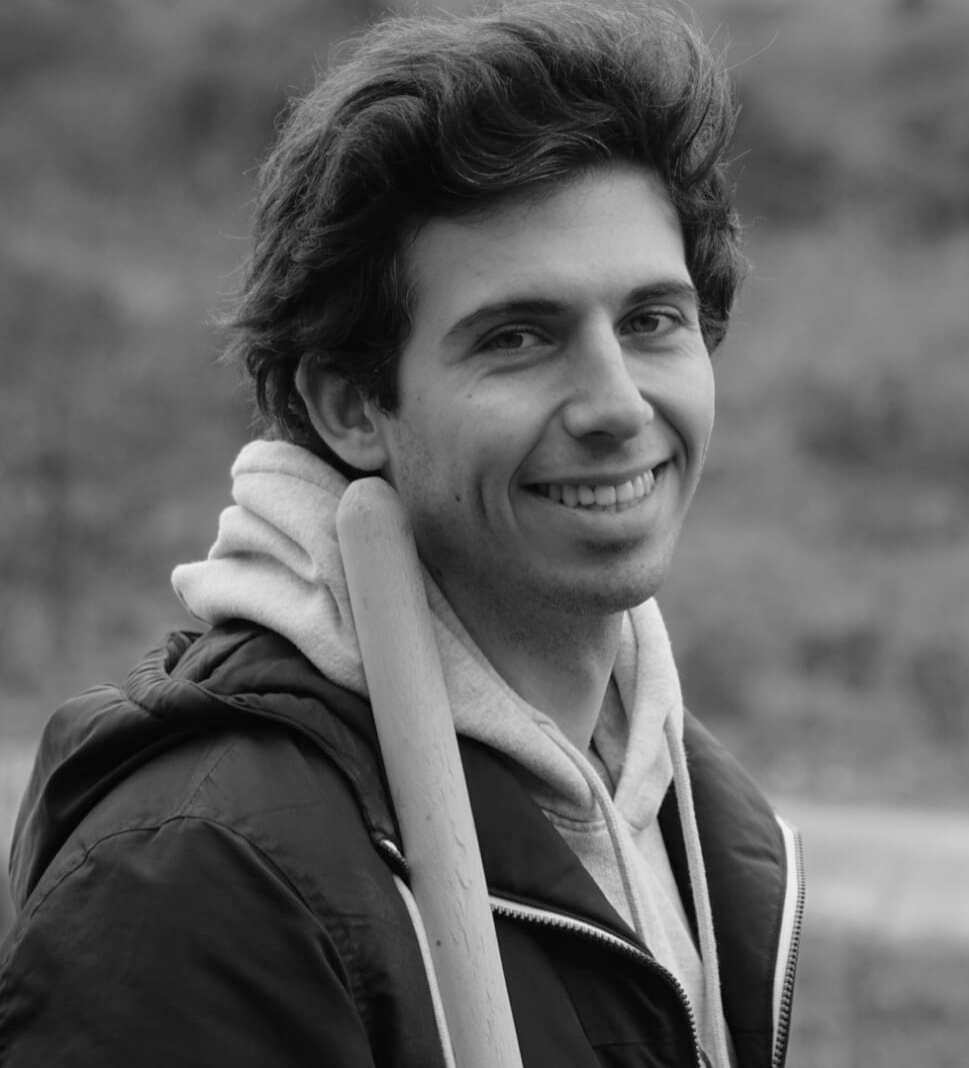}}]{Erhan Karakoca}
(Student Member, IEEE) completed his B.Sc. (with high Hons.) in electronics and communication engineering at Y{\i}ld{\i}z Technical University in Istanbul, Turkey. He is currently pursuing an M.Sc. in telecommunications engineering at Istanbul Technical University in Istanbul, Turkey. Also, he is a researcher in the Hisar Lab of T\"UB\.ITAK B\.ILGEM. Next-generation wireless communication systems, terahertz wireless communications, Bayesian statistics, and machine learning are among his research interests.
\end{IEEEbiography}

\vspace{11pt}

\begin{IEEEbiography}[{\includegraphics[width=1.1in,height=1.1in,keepaspectratio]{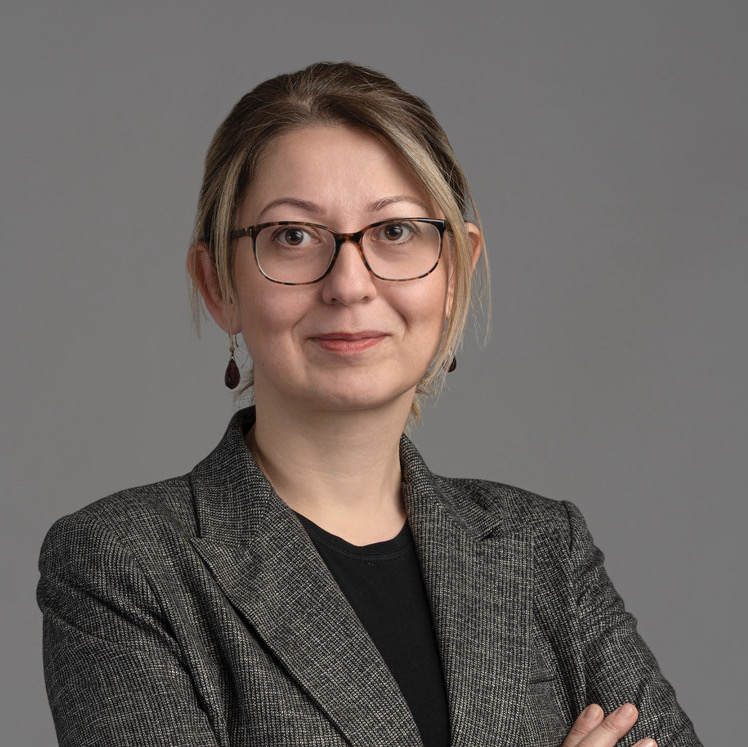}}]{G\"{u}ne\c{s} Karabulut Kurt}
 (Senior Member, IEEE) is currently an Associate Professor of Electrical Engineering at Polytechnique Montr\'eal, Montr\'eal, QC, Canada.  She received the B.S. degree with high honors in electronics and electrical engineering from the Bogazici University, Istanbul, Turkey, in 2000 and the M.A.Sc. and the Ph.D. degrees in electrical engineering from the University of Ottawa, ON, Canada, in 2002 and 2006, respectively. From 2000 to 2005, she was a Research Assistant at the University of Ottawa. Between 2005 and 2006, Gunes was with TenXc Wireless, Canada. From 2006 to 2008, she was with Edgewater Computer Systems Inc., Canada. From 2008 to 2010, she was with Turkcell Research and Development Applied Research and Technology, Istanbul. Gunes was with Istanbul Technical University between 2010 and 2021. She is a Marie Curie Fellow and has received the Turkish Academy of Sciences Outstanding Young Scientist (T\"UBA-GEBIP) Award in 2019. She is an Adjunct Research Professor at Carleton University. She is also currently serving as an Associate Technical Editor (ATE) of the IEEE Communications Magazine and a member of the IEEE WCNC Steering Board.
\end{IEEEbiography}

\vspace{11pt}

\begin{IEEEbiographynophoto}{Ali G\"{o}r\c{c}in}
(Senior Member, IEEE) received the B.Sc. degree in electronics and telecommunications engineering and the master's degree in defense
technologies from Istanbul Technical University, and the Ph.D. degree in
wireless communications from the University of South Florida (USF). After
working at Turkish Science Foundation (T\"UB\.ITAK) on avionics projects
for more than six years, he moved to the U.S. to pursue Ph.D. degree.
He worked with Anritsu Company during his tenure with USF and worked
with Reverb Networks and Viavi Solutions after his graduation. He is currently an Assistant Professor at Y{\i}ld{\i}z Technical University, Istanbul. He is
also the President of T\"UB\.ITAK B\.ILGEM.
\end{IEEEbiographynophoto}

\vfill

\end{document}